\documentclass[12pt]{article}

\usepackage{mathptmx}

\usepackage{amsmath,amssymb}

\newtheorem{lemma}{Lemma}[section]

\newtheorem{theorem}{Theorem}[section]
\newtheorem{proposition}{Proposition}[section]

\numberwithin{equation}{section}

\newcommand{\proof}{ $\triangleright$\quad}
\newcommand{\qed}{\hfill $\triangleleft$}

\begin{document}

\title {Quantum energy-mass spectrum of Yang-Mills bosons}
 
\author{Alexander  Dynin\\
\textit{\small Department of Mathematics, Ohio State University}\\
\textit{\small Columbus, OH 43210, USA}, \texttt{\small dynin@math.ohio-state.edu}}

\date{}

\maketitle

\begin{abstract}
A non-perturbative  quantization  of  the Yang-Mills energy-mass functional  with a compact  semi-simple gauge  group entails  an infinite discrete  energy-mass  spectrum of  gauge bosons.  The bosonic  spectrum is bounded from below, and has a positive mass gap due to the quartic self-interaction term of pure Yang-Mills Lagrangian (with no  Higgs term   involved).  This quantization   is based on  infinite-dimensional analysis  in Kree nuclear triple of  sesqui-holomorphic functionals of initial data for the  the  non-linear classical Yang-Mills equations in the temporal gauge.


 \medskip
 AMS Subject Clasification: 35S99, 81T13, 81T16

\end{abstract}

\section{Introduction}
\subsection{Physical background}
A mass gap  for the subnuclear weak and strong forces is suggested by  experiments  in accordance with  Yukawa principle:  \emph{A  limited force range  suggests a massive carrier}. 

As this paper shows,   a positive  mass gap for  a \emph{pure} quantum     relativistic  Yang-Mills theory (with any semi-simple gauge group) can be  mathematically deduced  from the \emph{quartic} self-interaction term in the pure relativistic Yang-Mills Lagrangian with no Higgs term. 
 
\medskip
The   energy-mass  time component  $P^0$ of the classical  Noether energy-momentum relativistic vector $P^\mu$  is not a Poincare  scalar. I work  only  in the \emph{rest Lorentz frames} where  the  energy-momentum vector is a direction vector of   the time axis.The Yang-Mills energy-momentum 4-vector is time-like (see \textsc{glassey-strauss} \cite{Glassey}).  
 
   Any  Poincare frame is relativistically equivalent to a  Yang-Mills rest Lorentz frame, and any  Yang-Mills gauge is equivalent to the  temporal gauge (see, e.g, \textsc{faddeev-slavnov} \cite[Chapter III, Section 2]{Faddeev}. Since the quantization is invariant with respect to the residual Poincare and gauge symmetries, the bosonic spectrum and  its  spectral  gap are \emph{relative}  Poincare and gauge quantum invariants.

The paper is a formulation and a proof of the  spectral Theorem 4.1. This is done  in four steps: 
\begin{description}
    \item[A] \ In the temporal gauge,   Yang-Mills fields (i.e. solutions of \emph{relativistic} Yang-Mills equations) are in one-one correspondence  with their  constrained  Cauchy data (see, e.g.  \textsc{goganov-kapitanskii} \cite{Goganov}).  
    
    This parametrization  of the classical Yang-Mills fields is  advantageous in three ways: 
    \begin{itemize}
  \item  Cauchy data carry a \emph{positive definite} scalar product.  
  \item  The non-linear constraint equation for Cauchy data is \emph{elliptic}.
   \item Yang-Mills energy-mass functional of Cauchy data is \emph{not} a relativistic  conformal invariant.
\end{itemize}   
The elliptic equation is solved  via a gauge version of  classical Helmholtz decomposition of vector fields.  The solution provides a \emph{linearization of  non-linear  constraint  manifold} of Cauchy data.
    
\item[B]\  In the line of  I. Segal  non-linear quantization program   (see, \textsc{segal} \cite{Segal-60}) along with  the quantum postulate  of  \textsc{bogoliubov-shirkov} \cite[Chapter II]{Bogoliubov}) I quantize   the  rest energy-mass functional of  Cauchy data as a  tame operator in a Kree nuclear triple of \cite{Kree-1}, \cite{Kree-2} (see also \textsc{meyer}\cite{Meyer}). Note that according to  the quantum postulate  no operators fields are involved.
The  quantization is chosen to be \emph{anti-normal} (aka anti-Wick,  Berezin,  or diagonal quantization).

  \item[C] \ I extend  \textsc{agarval-wolf} \cite{Agarwal} symbolic calculus to   show that the  Weyl  symbol of  the anti-normal energy-mass operator contains a \emph{positive quadratic mass term } which is absent in the classical  energy-mass  functional. \footnote{For other   applications of the symbolic calculus to quantum  dynamics  see \textsc{dynin} \cite{Dynin-10}.}

  \item[D]\     It follows that the  anti-normal energy-mass operator dominates  a shift of the number operator. Splitting off the  invariant spaces that are irreducible under bosonic permutations, I show that the corresponding bosonic spectrum is \emph{infinite and discrete}.
   \end{description}
   
\emph{All new defined terms in the text are introduced via emphasizing in italics.
The beginning and the end  of a proof are marked  by $\triangleright$ and $\triangleleft$ .}

\section{Yang-Mills  energy-mass functional}
\subsection{Gauge groups}

The \emph{global gauge  group}   $\mathbb{G}$ of a  Yang-Mills theory is   a  
connected semi-simple compact Lie group with the  Lie algebra $\mbox{Ad}(\mathbb{G})$. 

The notation $\mbox{Ad}(\mathbb{G})$ indicates that the Lie algebra carries the adjoint representation $\mbox{Ad}(g)X=gXg^{-1}, g\in\mathbb{G}, a\in Ad(\mathbb{G})$, of the group $\mathbb{G}$ and the corresponding self-representation $\mbox{ad}(X)Y=[X,Y],\ X,Y\in\mbox{Ad}(\mathbb{G})$.  Then $\mbox{Ad}(\mathbb{G})$ is identified with a Lie algebra of skew-symmetric matrices and the matrix commutator as Lie bracket with  the \emph{positive  definite}  Ad-invariant  scalar  product
\begin{equation}
\label{eq:scalar}
X\cdot Y \ \equiv\  \mbox{Trace}(X^TY), 
\end{equation}
where $X^T=-X$ denotes the matrix transposition  (see, e.g.  \textsc{zhelobenko} \cite[section 95]{Zhelobenko}).

\bigskip
Let the Minkowski space $\mathbb{M}$ be oriented and time oriented with  the Minkowski metric signature $(+,-,-,-)$. In a Minkowski coordinate systems 
$x^\mu, \mu=0,1,2,3$,  the metric tensor is diagonal.
  In  the natural unit system, the time coordinate $x^0=t$. Thus  $(x^\mu)=(t,x^i),\  i=1,\ 2,\ 3$. 
  
 The \emph{local gauge  group} $\mathcal{G}$ is the group of  infinitely differentiable $\mathbb{G}$-valued functions
 $g(x)$ on $\mathbb{M}$ with the pointwise group multiplication.  The  \emph{local gauge Lie algebra}  $\mbox{Ad}(\mathcal{G})$ consists of  infinitely differentiable $\mbox{Ad}(\mathbb{G})$-valued functions   on $\mathbb{M}$ with the pointwise Lie bracket.   
 
$\mathcal{G}$ acts via the pointwise adjoint action on $\mbox{Ad}(\mathcal{G})$   and correspondingly on  $\mathcal{A}$, the real vector space of \emph{gauge   fields}   $A=A_\mu(x)\in\mbox{Ad}(\mathcal{G})$. 

 \smallskip
   Gauge fields $A$ define   the \emph{covariant partial derivatives}  
   \begin{equation}\label{}
  \partial_{A\mu}X\ \equiv\  \partial_\mu X- \mbox{ad}( A_\mu)X,\quad
X\in\mbox{Ad}(\mathcal{G}).  
\end{equation}

Any $g\in\mathcal{G}$ defines the affine \emph{gauge transformation}  
\begin{equation}\label{}
A_\mu\mapsto A_\mu^{g}\equiv\mbox{Ad}(g)A_\mu-(\partial_\mu g)g^{-1},\ A\in \mathcal{A},
\end{equation}
so that $A^{g_1}A^{g_2}=A^{g_1g_2}$.

\subsection{Yang-Mills fields}
Yang-Mills \emph{curvature tensor} $F(A)$ is the 
antisymmetric tensor
\begin{equation}\label{}
F(A)_{\mu\nu}\equiv\partial_\mu A_\nu-\partial_\nu A_\mu-[A_\mu,A_\nu].
\end{equation} 
The curvature is gauge covariant:
 \begin{equation}\label{}
\partial_{A\mu }\mbox{Ad}(g)\ =\ \mbox{Ad}(g)
\partial_{A\mu},\quad
\mbox{Ad}(g)F(A)\ =\ F(A^{g}).
 \end{equation}
 
The \emph{Yang-Mills Lagrangian} (with the coupling constant   set to 1)
 \begin{equation}\label{}
 \label{eq:Lag}
  L= - (1/4)F(A)^{\mu\nu}\cdot F(A)_{\mu\nu}
  \end{equation}
  is invariant  under gauge transformations.

The corresponding Euler-Lagrange equation is a   2nd order non-linear  partial differential equation $\partial_{A\mu}F(A)^{\mu\nu} =0$, called 
 the \emph{Yang-Mills equation} 
\begin{equation}
\label{eq:YM}
 \partial_\mu F^{\mu\nu}\ +\ [A_\mu, F^{\mu\nu}]\ =\ 0.
\end{equation}
The solutions   $A$ are  \emph{Yang-Mills fields}. They  form the \emph{on-shell space} $\mathcal{M}$ of  the classical Yang-Mills theory. 

\medskip
\emph{From now on we assume  that  all space derivatives of   gauge  fields $A=A(t,x^k)$  vanish faster than any power of $x^kx_k$  as $x^kx_k\rightarrow \infty$, uniformly with respect to bounded $t$}. (This condition does not depend on  a Lorentz coordinate system.) Let $\mbox{Ad}\mathcal{G}$ denote the local Lie algebra of such gauge fields and 
$\mathcal{G}$ denote the corresponding infinite dimensional local Lie group.

\smallskip
In a Lorentz coordinate system we have the following  matrix-valued  time-dependent fields on  $\mathbb{R}^3$: 
\begin{description}
  \item Gauged electric vector field  $E(A)\equiv(F_{01},F_{02},F_{02})$,
  \item Gauged magnetic pseudo vector field  $B(A)\equiv(F_{23},F_{31},F_{12})$.
  \end{description}
Now the (non-trivial) energy-mass    conservation law   is  that   the time component 
  \begin{equation}
\label{eq:energy}
P^0(A)\equiv\int\!d^3x\, (1/2)(E^i\cdot E_i  + B^i\cdot B_i)
\end{equation}
of the relativistic Noether  energy-momentum vector is constant on-shell. Appropriately, $P^0(A)$
 has  the mass dimension.

At the same time,  by Glassey-Strauss Theorem \cite{Glassey}, the \emph{energy-mass density}  $(1/2)(E^i\cdot E_i  + B^i\cdot B_i)$    scatters asymptotically along the light cone as $t\rightarrow \infty$. This is a mathematical reformulation of the  physicists assertion that 
Yang-Mills fields propagate with the light velocity.

\subsection{First order formalism}
Rewrite  the    2nd order Yang-Mills equations (\ref{eq:YM})   in the temporal gauge 
$A_0(t,x^k)=0$ as the 1st order systems   of the \emph{evolution equations}  for the time-dependent $A_j(t,x^k)$,  $E_j(t,x^k)$ on  $\mathbb{R}^3$ as
\begin{equation}
\label{eq:evolution}
\partial_t A_k\  = \  E_k ,  \quad
\partial_tE_k \  = \  \partial_jF^j_k - [A_j,F^j_k],\ \quad\ F^j_k\ =\ \partial^j A_k - \partial_k A^j - [A^j,A_k].
\end{equation}
and the \emph{constraint  equations}
\begin{equation}
\label{eq:constraint}
 [A^k,E_k]    \ =\  \partial^kE_k, \quad \mbox{i.e.}\ \quad \partial_{k,A}E_k\ =\ 0\\
\end{equation}
By  \textsc{goganov-kapitanskii} \cite{Goganov}, the evolution system is  a semilinear first order  partial differential  system  with  \emph{finite speed propagation} of the initial data, and the  Cauchy  problem for it with  initial data at $t=0$ \begin{equation}
\label{ }
a(x_k)\ \equiv\ A(0,x_k), \quad \ e(x_k)\ \equiv\ E(0,x_k)
\end{equation}
is \emph{globally and uniquely solvable on} the whole Minkowski space $\mathbb{M}$.

Actually,  this was proved in \cite{Goganov} without any restriction on Cauchy data at the  infinity.

\smallskip

If the  constraint equations are satisfied  at $t=0$, then, in view of the evolution system, they are satisfied   for  all $t$ automatically. Thus the  \emph{1st order evolution system along with the  constraint equations for Cauchy data is equivalent  to the 2nd order Yang-Mills system}. Moreover the constraint equations are invariant under  \emph{time independent} gauge transformations. As the bottom line, we have
\begin{proposition}
\label{pro:trans}
In the temporal gauge Yang-Mills fields $A$ are in one-one correspondence with their  gauge transversal Cauchy data $(a,e)$ satisfying   the  equation $\partial_a e=0$.  
\end{proposition}

Let $\mathcal{A}^0=\mathcal{A}^0(\mathbb{R}^3)$ denote the  \emph{real} $\mathcal{L}^2$-space of Cauchy gauge vector fields  $a$ on $\mathbb{R}^3$.  
The associated   Sobolev-Hilbert spaces  (see, e.g.  \textsc{shubin} \cite[Section 25]{Shubin}) are  denoted   $\mathcal{A}^s,\ s\in\mathbb{R}$. The  intersection  $\mathcal{A}^\infty_0\equiv\bigcap_s\mathcal{A}^s$ with the projective limit topology is a nuclear real Frechet space of smooth $a$.  The  union $\mathcal{A}^{-infty}\equiv\bigcup_s \mathcal{A}^{-s}$ with the inductive limit topology is its dualspace of $\mathcal{A}^\infty_0$.

\smallskip
Let $\mathcal{G}^s,\ s>3/2,$ be the infinite dimensional Frechet Lie groups with the  Lie algebras $\mathcal{A}^s\ s>3/2$.

The intersection $\mathcal{G}^\infty\equiv\bigcap_s\mathcal{G}^s$
is an infinite dimensional Lie group with the \emph{nuclear} Lie algebra
$\mathcal{A}^\infty$. The local gauge transformations $a^g$ by 
$g\in\mathcal{G}^\infty$ define  continuous  left action $\mathcal{G}^\infty\times\mathcal{A}^s\rightarrow
\mathcal{A}^{s-1}$.

  Local gauge transformations 
\begin{equation} \label{eq:gauge}
a_k^g=\mbox{Ad}(g)a_k-(\partial_k g)g^{-1},
\quad g\in\mathcal{G}^\infty,\ a\in \mathcal{A}^s,
\end{equation} 
define continuous  left action of $\mathcal{G}^s$ on $\mathcal{A}^s$.

The Sobolev-Hilbert  spaces  $\mathcal{S}^s$  of  smooth   Cauchy gauge  electric fields $e$ on $\mathbb{R}^3$ with the corresponding action $e^g$  of the local gauge group  $\mathcal{G}^\infty$ are defined the same way.

\subsection{Gauged vector calculus}

Let $\mathcal{U}^s$  denote the Sobolev-Hilbert spaces $\mbox{Ad}(\mathbb{G})$-valued functions $u$ on $\mathbb{R}^3$.

Consider the continuous  vector calculus  operators gauged by $a\in\mathcal{A}^{\infty}$

\smallskip
\emph{Gauged gradient}
 \begin{equation}
\label{ }
  \mbox{grad}_{a}:\ \mathcal{U}^{s}\rightarrow \mathcal{S}^{s-1},\quad\ \mbox{grad}_{a}u \ \equiv  \partial_k u - [a_k,u],
\end{equation} 
\indent\emph{Gauged divergence}
\begin{equation}
\label{ }
  \mbox{div}_{a}:\ \mathcal{S}^{s}\rightarrow \mathcal{U}^{s-1},\quad\ 
  \mbox{div}_{a}e \ \equiv  \partial_k e_k - [a_k,e_k],
\end{equation} 
\indent\emph{Gauged Laplacian}
\begin{equation}
\label{ }
  \triangle_a :\ \mathcal{U}^{s}\rightarrow \mathcal{U}^{s-2},\quad\  
  \triangle_a \ \equiv\  \mbox{div}_{a}u\,\mbox{grad}_{a}u,
\end{equation} 
 The 1st order partial differential operators $-\mbox{grad}_{a}$ and  $\mbox{div}_{a}$  are adjoint  with respect to the $\mathcal{L}^2$ scalar product:
\begin{equation}\label{adjoint}
\langle\  -\mbox{grad}_{a} u\ |\ v\ \rangle\ =\  
\langle\, u\ |\ \mbox{div}_{a}v\, \rangle.\end{equation}
The  gauged Laplacian $\triangle_a$ is a  2nd order partial differential operator. Since its principal part is the usual Laplacian $\triangle$, the operator $\triangle_a$ is elliptic. \begin{proposition}
\label{pro:Lap}
The gauge Laplacian  $\triangle_a$ is  an invertible operator  from 
$\mathcal{U}^{s+2}$ onto $\mathcal{U}^{s}$ for all $s\geq 0$.
\end{proposition}
\begin{lemma}\label{lemma:inj}
$\triangle_a u=0,\ u\in\mathcal{U}^1_0,$ if and only if  $u=0$.
\end{lemma}  
\proof
 $u\cdot [a,u]=-\mbox{Trace}(uau-uua)=0$ so that
\begin{equation}
\label{ } 
u\cdot \mbox{grad}_{a}\,u \ =
u\cdot\mbox{grad}\, u \ =\  
(1/2)\mbox{grad}\,(u\cdot u)\ =\ 0.
\end{equation}
This shows that   for $u\in\mathcal{U}^1_0$ we have $\mbox{grad}_a u =0$    if and only if  $u=0$. \qed

Next, by the equality (\ref{adjoint}),
\begin{equation}\label{}
\langle\, \triangle_a u\ |\ u\, \rangle\ =\
\langle\  -\mbox{grad}_{a}u\ |\ \mbox{grad}_{a}u\ \rangle,\   u\in\mathcal{U}^1_0.
\end{equation}
Thus   $\triangle_a u=0,\ u\in\mathcal{U}^1_0,$ if and only if  $u=0$. \qed

\smallskip
Both Laplacian  $\triangle$ and gauge Laplacian    $\triangle_a$ map 
$\mathcal{U}^{s+2}$ into $\mathcal{U}^s$.

The Laplace operator 
 is invertible from  $\mathcal{U}^{s+2}$ onto $\mathcal{U}^{s}$ whatever $s\geq 0$ is.
Since $\triangle - \triangle_a$ is a 1st order differential operator, the operator 
 $\triangle_a:\mathcal{U}^{s+2}\rightarrow\mathcal{U}^s$ is a Fredholm operator of zero index. Then, by   Lemma \ref{lemma:inj}, the inverse 
$\triangle_{a}^{-1}:\ \mathcal{U}^{s}\rightarrow\mathcal{U}^{s+2}$ exists
for all $s\geq 0$. \qed

\medskip
Proposition \ref{pro:Lap} shows that the operator  $\mbox{div}_a: \mathcal{U}^s\rightarrow\mathcal{U}^{s-1}$ is surjective and
 the operator  $\mbox{grad}_a: \mathcal{U}^s\rightarrow\mathcal{U}^{s-1}$ is injective. 

Let
\begin{equation}\label{Helm}
\Pi_a\ \equiv\ 
 \mbox{grad}_a\triangle_{a}^{-1}\mbox{div}_{a}
\end{equation}
Both $\Pi_a$ and $\mathbf{1}-\Pi_a$ are  pseudodifferential operators of order 0, and,
therefore are $\mathcal{L}^2$- bounded.

By computation, 
\begin{equation}
\label{ }
\Pi_a^\dag=\Pi_a,\quad \Pi_a^2=\Pi_a,\quad \Pi_a\mbox{grad}_{a}=\mbox{grad}_{a},\quad 
\mbox{div}_a(\mathbf{1}-\Pi_a)=0. 
\end{equation}
Therefore $\Pi_a$ is an  $\mathcal{L}^2$-orthogonal projector of $\mathcal{U}^s$ onto  the space of gauge longitudinal  vector fields, i.e. the range   of the operator 
$\mbox{grad}_a:\ \mathcal{U}^{s+1}\rightarrow \mathcal{U}^s$; and the operator $\mathbf{1}-\Pi_a$ is an $\mathcal{L}^2$ bounded projector of $\mathcal{U}^s$ onto  the space of gauge transversal vector fields, i.e. the null space  of the operator 
$\mbox{div}_{a}: \mathcal{U}^s\rightarrow\mathcal{U}^{s-1}$.

Now Proposition \ref{pro:trans} implies 
 \begin{proposition}
\label{pro:proj}
In the temporal gauge Yang-Mills fields $A$ are in one-one correspondence with the vector bundle of 
the ranges of projectors $\Pi_a$ over the base $\mathcal{A}^\infty$
\end{proposition}

\subsection{Gelfand-Schwartz  triple of constrained Cauchy data}
 
Let  $\mathcal{T}^\infty_a\subset\mathcal{S}^\infty$  denote  the nuclear  Frechet  space  of gauge transversal  gauge electric vector fields $e_a\equiv \ e-\Pi_a(e)$, and $\mathcal{T}^0$ be its completion in  $\mathcal{S}^0$.

The family of orthogonal projectors  $a\mapsto \Pi_a$ is a continuous mapping of  
 $\mathcal{A}^\infty$ to the algebra of bounded operators on $\mathcal{S}^0$.  Since for $a$ sufficiently close to $a_o$ the  operators $1-\Pi_a+\Pi_{a_o}$ are invertible and
$\Pi_a\Pi_{a_o}=\Pi_a(1-\Pi_a+\Pi_{a_o})\Pi_{a_o}$, the continuous mappings $\Pi_a\Pi_{a_o}:  \Pi_{a_o}(\mathcal{S}^0)\rightarrow  \Pi_{a}(\mathcal{S}^0)$ are invertible. Thus  the  vector  bundle $\mathcal{T}^0$ of the gauge transversal  spaces 
$\mathcal{T}^0_a$ is a locally trivial real vector bundle over $\mathcal{A}^\infty$.
  
Gauge invariance of the constraint manifold of Cauchy data under the (residual) gauge group  implies the gauge covariance of projectors $\mathbf{1}-\Pi_a$, and so of  the  bundles.
Since   a  Hilbert  bundle structure group is   smoothly contractible  (see \textsc{kuiper} \cite{Kuiper}), the bundle $\mathcal{T}^0$
is isomorphic to the trivial gauge covariant Hilbert space bundle over its base:  an isomorphism is defined by a smooth family of orthonormal bases of the bundle fibers. All such trivialisations intertwine with the action of the residual gauge group.
They define linearly isomorphic global Hilbert  coordinate charts on the constraint Cauchy data manifold $\mathcal{C}^0\ \cong \mathcal{A}^0\times \mathcal{T}^0_0$.  

\medskip
From the  $N$-representation  of nuclear Schwartz $\mathcal{S}$-spaces (cp., e.g. \textsc{reed-simon} \cite[Theorem V.13]{Reed} on the fibers, we get a locally trivial  bundle of  nuclear Schwartz $\mathcal{S}$-spaces  of constrained $e$-fields over the same base  along with    the natural Gelfand nuclear triple of real topological vector spaces
\begin{equation}\label{eq:Gelfand}
\mathcal{C}:\ \mathcal{C}^\infty\  \equiv \mathcal{A}^\infty_0\times 
\mathcal{T}^\infty_0\subset\ \mathcal{C}^0\ \equiv \mathcal{A}^0\times 
\mathcal{T}^0\subset\ \mathcal{C}^{-\infty}\ \equiv \mathcal{A}_0^{-\infty}\times 
\mathcal{T}_0^{-\infty}.
\end{equation}
where $\mathcal{C}^\infty$  is a nuclear Frechet space of smooth $(a,e^o)$, and 
$\mathcal{C}_0^{-\infty}$ is  the  dual of $\mathcal{C}^\infty_0$,  with the duality defined by the inner product in $\mathcal{C}^0$.

\medskip
The assignment $(a,e^o)\mapsto z=(1/\sqrt{2})(a+ie^o)$ converts the real  Gelfand triple (\ref{eq:Gelfand})
into a complex  Gelfand triple, so that $\Re\mathcal{C}\ \equiv\mathcal{A}$ and 
$\Im\mathcal{C}\ \equiv\mathcal{T}_o$ are its real and imaginary parts.

 The  complex conjugation
\begin{equation}
\label{ }
z^*\ =\  (1/\sqrt{2})(a+ie^o)^*\ \equiv\ (1/\sqrt{2})(a-ie^o),\ z\mapsto z^*:\ \mathcal{C}\rightarrow
\mathcal{C}^{-\infty}
\end{equation} 
The  (anti-linear on the left and linear on the right) Hermitian product is  defined  on   $\mathcal{C}^0$ as
\begin{equation}
\label{}
z^*z\ \equiv\ (1/2)\int\!d^3 x\ (a\cdot a + e^o\cdot e^o)(x)
\end{equation}
 extended to the anti-duality between $\mathcal{C}^\infty$ and  $\mathcal{C}^{-\infty}$. Accordingly,  \emph{the notation $z$ is reserved for the elements of the  space $\mathcal{C}^\infty$, and  the notation $z^*$ for the elements of the space $\mathcal{C}^{-\infty}$}.
The result is
\begin{proposition}
The constraint  Cauchy manifold is a complex nuclear Gelfand-Schwartz triple.
 \end{proposition}

\section{Quantization}
\subsection{Review of \textsc{kree} \cite{Kree-1},  \cite{Kree-2}}
The  triple $\mathcal{C}$ is a complex nuclear triple with the Hermitian conjugation $*$.   By  \textsc{kree} \cite{Kree-1} and \cite{Kree-2} (also \textsc{meyer}\cite{Meyer}),  there is the associated Kree nuclear triple with the induced  Hermitian conjugation $*$
\begin{equation}
\label{eq:Kree}
 \mbox{Exp}(\mathcal{C}^{-\infty})\ \subset\ \mathcal{B}(\mathcal{C}^{-\infty})\ \subset\ \mbox{Ent}(\mathcal{C}^{\infty})
\end{equation}
where 
\begin{itemize}
 \item  $\mbox{Ent}(\mathcal{C}^{\infty})$ is the complete nuclear space of all entire holomorphic functionals on $\mathcal{C}^{\infty}$ with the topology of compact convergence.

  \item $\mathcal{B}(\mathcal{C}^{-\infty})$ is the   Bargmann subspace of square integrable entire holomorphic functionals  on $\mathcal{C}^{-\infty}$ with respect to 
 the Gauss-Minlos probability measure $\gamma$. This is a complete Hilbert space  identified with its $*$-dual with by the  Hermitian form
  \begin{equation}
\label{eq:form}
 \langle\ \Psi\ |\ \Phi\ \rangle\ \equiv\ \int \!d\gamma\: \Psi^*(z)\Phi(z^*),  
 \end{equation}
where the $*$-dual $\Psi^*(z$ is the complex conjugate of $\Psi(z^*)$.

  \item $\mbox{Exp}(\mathcal{C}^{-\infty})$ is the nuclear space of entire functionals  of exponential type on $\mathcal{C}^{-\infty}$ identified with the   $*$-dual of 
   $\mbox{Ent}(\mathcal{C}^{\infty})^*$ via the $*$-duality  induced by (\ref{eq:form}). 
   
  \item  The \emph{exponential states}, aka coherent states, $e^\zeta\in\mbox{Exp}(\mathcal{C}^{-\infty}),\ \zeta\in \mathcal{C}^{\infty}$,  
   \begin{equation}
\label{ }
e^\zeta(z^*)\ \equiv e^{z^*\zeta},\quad z^*\in \mathcal{C}^{-\infty}, \quad
 \langle\ e^\zeta\ |\ e^w\ \rangle\ =\ e^{\zeta^*w},
\end{equation}
belong to $\mbox{Exp}(\mathcal{C}^{-\infty})$. They form the  \emph{continual exponential basis} of   $\mbox{Ent}(\mathcal{C}^{\infty})$,  with the \emph{continual  coordinates}
   \begin{equation}
   \label{eq:basis}
\tilde{\Psi}(\zeta)\ \equiv   \langle\ \Psi^*\ |\ e^\zeta\ \rangle, \quad   \Psi\in 
\mbox{Ent}(\mathcal{C}^{\infty}).
\end{equation}
This   \emph{complex Fourier transform} (aka Borel transform) is a topological $*$-automorhism of the Kree triple:
For $\Psi\in \mbox{Ent}(\mathcal{C}^{\infty}),\ \Phi\in \mbox{Exp}(\mathcal{C}^{\infty})$
 \begin{equation}
\label{eq:Planch}
 \langle\ \Psi^*(\zeta)\ |\ \Phi(\zeta^*)\ \rangle\
 =\ \langle\ \Psi^*(z)\ |\ \Phi(z^*)\ \rangle.
\end{equation}
\end{itemize}
By Grothendieck kernel theory, the nuclearity of the  Kree triple implies that the locally convex vector spaces  $\mathcal{O}\big(\cdot\rightarrow\cdot\big)$ of continuous linear operators   are topologically isomorphic to the complete sesqui-linear  tensor products (both spaces are endowed with the topology of compact uniform convergence).
\begin{eqnarray}
\label{eq:general} \mathcal{O}\big(\mbox{Exp}(\mathcal{C}^{-\infty}) \rightarrow\mbox{Ent}(\mathcal{C}^\infty) \big)
 & \simeq &  \mbox{Ent}(\mathcal{C}^{\infty})^*\widehat{\otimes}\: \mbox{Ent}(\mathcal{C}^\infty),
 \\
\label{eq:tame} \mathcal{O}\big(\mbox{Exp}(\mathcal{C}^{-\infty}) \rightarrow\mbox{Exp}(\mathcal{C}^{-\infty})  \big)
 & \simeq &  \mbox{Ent}(\mathcal{C}^{\infty})^*\widehat{\otimes}\:\mbox{Exp}(\mathcal{C}^{-\infty}), 
 \\
\label{eq:smooth}  \mathcal{O}\big(\mbox{Ent}(\mathcal{C}^\infty) \rightarrow \mbox{Exp}(\mathcal{C}^{-\infty})\big)
 & \simeq &   \mbox{Exp}(\mathcal{C}^{-\infty})^*\widehat{\otimes}\:  \mbox{Exp}(\mathcal{C}^{-\infty}) 
\end{eqnarray}
where the  operators are \emph{tame} in the case of (\ref{eq:tame}) and   \emph{smoothing} in the case of (\ref{eq:smooth}).
 
The formulas present the one-to-one correspondence  between operators and the  sesqui-holomorphic kernels of their matrix elements. 

The nuclear Gelfand-Schwartz  triple of the sesqui-Hermitian direct products   
\begin{equation}
\label{ }
(\mathcal{C}^{-\infty})^*\times \mathcal{C}^{-\infty}\ \subset\ 
(\mathcal{C}^{0})^*\times \mathcal{C}^{0}\ \subset\
(\mathcal{C}^{\infty})^*\times \mathcal{C}^{\infty}
\end{equation}
carries the Hermitian conjugation
\begin{equation}
\label{ }
(z^*,w)^* \equiv\ (w^*,z)
\end{equation} 
The associated Kree triple of  sesqui-holomorphic kernels consists of
\begin{eqnarray}
\label{eq:kernels}
 \mbox{Exp}\big((\mathcal{C}^{-\infty})^*\times\mathcal{C}^{-\infty}\big) & \simeq & 
  \mbox{Exp}(\mathcal{C}^{-\infty})^*\widehat{\otimes}\:  \mbox{Exp}(\mathcal{C}^{-\infty}), \\
    \mathcal{B}\big((\mathcal{C}^{-\infty})^*\times\mathcal{C}^{-\infty}\big) & \simeq & 
   \mathcal{B}(\mathcal{C}^{-\infty})^*)\widehat{\otimes}\:  \ \mathcal{B}(\mathcal{C}^{-\infty}), \\
  \mbox{Ent}\big((\mathcal{C}^{\infty})^*\times\mathcal{C}^\infty\big) & \simeq & 
   \mbox{Ent}(\mathcal{C}^{\infty})^*\widehat{\otimes}\: \mbox{Ent}(\mathcal{C}^\infty),
\end{eqnarray}
where  $ \mathcal{B}\big((\mathcal{C}^{-\infty})^*\times\mathcal{C}^{-\infty}\big)$
is the the Bargmann-Hilbert  space with respect to the direct  product of the probability Radon measures on $\mathcal{C}^{-\infty})^*\times\mathcal{C}^{-\infty}$.

The corresponding exponential functionals are 
\begin{equation}
\label{eq:biex}
e^{(\zeta^*,\eta)}\big((z^*,w)^*\big)\ =\ e^{w^*\eta+\zeta^*z}.
\end{equation}

\subsection{Operator symbols}
Kree triple  (\ref{eq:Kree}) carries the bosonic canonical linear  representation of   the triple	$\mathcal{C}$  by continuous linear transformations of $\zeta\in \mbox{Exp}(\mathcal{C}^{\infty})$ and $\zeta^*\in\mbox{Ent}(\mathcal{C}^{\infty})$ into  the adjoint operators of  \emph{creation and annihilation} continuous operators of multiplication and directional differentiation
\begin{eqnarray}
\hat{\zeta}:\  \mbox{Exp}(\mathcal{C}^{-\infty})\ \rightarrow  \mbox{Exp}(\mathcal{C}^{-\infty}),& \quad &
\hat{\zeta}\Psi(z^*) \equiv  (z^*\zeta)\Psi(z^*),\\
\widehat{\zeta^*}:\ \mbox{Ent}(\mathcal{C}^{\infty})\ \rightarrow 
\mbox{Ent}(\mathcal{C}^{\infty}), & \quad &
 \widehat{\zeta^*}\Psi(z) \equiv 
(\zeta^*z)\Psi(z), \\
\widehat{\zeta^*}^\dag:\ \mbox{Exp}(\mathcal{C}^{-\infty})\ \rightarrow \mbox{Exp}(\mathcal{C}^{-\infty}),& \quad &
\widehat{\zeta^*}^\dag\Psi(\zeta^*)  \equiv  \partial_{\zeta^*}\Psi(z^*),\\
\hat{\zeta}^\dag:\ \mbox{Ent}(\mathcal{C}^{\infty})\ \rightarrow 
\mbox{Ent}(\mathcal{C}^{\infty}),
& \quad &
 \hat{\zeta}^\dag\Psi(z) \equiv  \partial_\zeta\Psi(z).
\end{eqnarray} 
Their basic properties:
\begin{enumerate}
   \item 
 Bosonic commutation  relation
\begin{equation}
\label{eq:CCR}
[\widehat{\zeta^*}^\dag,\hat{\eta}]\ =\ \zeta^*\eta, \quad [\hat{\zeta}^\dag,\widehat{\eta^*}]\ =\ w^*\zeta.
\end{equation}

\item The exponentials $e^{\eta^*},\ \eta^*\in(\mathcal{C}^{-\infty})$  and 
$e^{\eta},\ \eta\in(\mathcal{C}^{\infty})$ are the eigenstates of the annihilation operators
\begin{equation}
\label{ }
\widehat{\zeta^*}^\dag e^\eta\ =\ (\zeta^*\eta)e^v,\quad \hat{\zeta}^\dag e^{\eta^*}\ =\ (\eta^*\zeta)e^{\eta^*}.
\end{equation}  
\end{enumerate}

Creators and annihilators generate strongly continuous abelian operator groups in 
$\mbox{Exp}(\mathcal{C}^{\infty})$ and $\mbox{Ent}(\mathcal{C}^{\infty})$ parametrized by $\zeta$ and $\zeta^*$:
\begin{eqnarray}
\label{eq:1}
e^{\hat{\zeta}}:\mbox{Exp} (\mathcal{C}^{-\infty})\rightarrow\mbox{Exp} (\mathcal{C}^{-\infty}),\quad & &
e^{\hat{\zeta}}\Psi(z^*)\ =\ e^{z^*\zeta}\ \Psi(z^*);\\
\label{eq:2}
e^{\hat{\zeta}^\dagger}:\mbox{Ent} (\mathcal{C}^{\infty})\rightarrow\mbox{Ent} (\mathcal{C}^{\infty}),\quad & &
e^{\hat{\zeta}^\dagger}\Psi(z)\ =\Psi(z+\zeta);\\
\label{eq:3}
e^{\widehat{\zeta^*}}:\mbox{Ent} (\mathcal{C}^{\infty})\rightarrow\mbox{Ent} (\mathcal{C}^{\infty}),\quad & &
e^{\widehat{\zeta^*}}\Psi(z)\ =\ e^{\zeta^*z}\ \Psi(z);\\
\label{eq:4}
e^{\widehat{\zeta^*}^\dagger}:\mbox{Exp} (\mathcal{C}^{-\infty})\rightarrow\mbox{Exp} (\mathcal{C}^{-\infty}),\quad & &
e^{\widehat{\zeta^*}^\dagger}\Psi(z^*)\ =\  \Psi(z^*+\zeta^*).
\end{eqnarray}

  \emph{Normal, Weyl,  anti-normal quantizations}   are  the operators $\widehat{M}$ with \emph{cokernels} $M(\theta^*,\eta)$ 
  \begin{equation}
\label{ }
\widehat{M}:\ \mbox{Exp}(\mathcal{C}^{\infty})\rightarrow \mbox{Ent}(\mathcal{C}^{\infty}),\quad
M\in\mbox{Ent}\big((\mathcal{C}^{\infty})^*\times\mathcal{C}^{\infty}\big)
\end{equation} 
  defined by their exponential matrix elements 
\begin{eqnarray}
\label{eq:n}
\langle\ e^{z^*}\ |\ \widehat{M}_n\ |\ e^{w}\ \rangle\ & \equiv & \big\langle \ M_n(\theta^*,\eta)\  \big|\ 
\langle e^{z^*} | e^{w}\ \rangle\ \big\rangle,
\\
\label{eq:w}
\langle\ e^{z^*}\ |\ \widehat{M}_w\ |\ e^{w}\ \rangle\ & \equiv & \big\langle\ M_w(\theta^*,\eta)\  \big|\ 
\langle e^{z^*}| e^{\hat{\theta}+\widehat{\eta^*}^\dagger}| e^{w}\ \rangle\ \big\rangle,
\\
\label{eq:an}
\langle\ e^{z^*}\ |\ \widehat{M}_{an}\ |\ e^{w}\ \rangle\ & \equiv & \big\langle\ M_{an}(\theta^*,\eta)\ \big|\ 
\langle e^{z^*}| e^{\widehat{\theta^*}^\dagger} e^{\hat{\eta}}| e^{w}\ \rangle\ \big\rangle.
\end{eqnarray}

Since
\begin{eqnarray}
& & 
\langle\ e^{z^*}\ | e^{\hat{\eta}}e^{\widehat{\theta^*}^\dagger}\ | e^{w}\ \rangle\  =\
\langle\  e^{\hat{\eta}^\dagger}e^{z^*}\ |\ \widehat{\theta^*}^\dagger e^{w}\ \rangle \\
 & &
=\ \langle\  e^{z^*\eta}e^{z^*}\ |\  e^{\theta^*z}e^{w}\ \rangle\ =\ e^{z^*\eta+\theta^*z}e^{z^*w},
\end{eqnarray}
one has  
\begin{equation}
\label{ }
\langle\ e^{z^*}\ |\ \widehat{M}_n\ |\ e^{w}\ \rangle\ =\ \langle\ M_n(\theta^*,\eta)\ |\  e^{z^*\eta+\theta^*z}\rangle e^{z^*w}\ =\ \tilde{M}_n(z^*,w)e^{z^*w}
\end{equation}
where $\tilde{M}_n(z^*,w)$ is the sesqui-linear Fourier transform of  $M_n(\theta^*,\eta)$ (see (\ref{eq:biex})). Thus $\tilde{M}_n(z^*,w)e^{z^*w}$ is the normal kernel of
$\widehat{M}_n$.

Since  $M_n(\theta^*,\eta)$  is an arbitrary co-kernel, by (\ref{eq:kernels}), any continuous  linear operator $Q$ from $\mbox{Exp}(\mathcal{C}^{\infty})$ to $\mbox{Ent}(\mathcal{C}^{\infty})$ 
has a unique normal  kernel $\tilde{M}_n^Q(z^*,w)$.

By the Taylor expansion centered at origin, the sesqui-entire functionals   are uniquely  defined  by   their restrictions   to the real diagonal $(z^*,w=z)$ so that  the \emph{normal symbol}
of the operator $Q$
\begin{equation}
\label{eq:nsymb}
\sigma^Q_n(z^*,z)\ \equiv e^{-z^*z}\tilde{M}_n^Q(z^*,z)
\end{equation}
exists and defines $Q$ uniquely.

\smallskip
By Baker-Campbell-Hausdorff commutator formula and the canonical commutation relations (\ref{eq:CCR}),
 \begin{equation}
\label{eq:BCH}
e^{\hat{\eta}}e^{\widehat{\theta^*}^\dagger}\ =\ e^{\hat{\eta}+\widehat{\theta^*}^\dag}
e^{\theta^*\eta/2}, \quad  e^{\widehat{\theta^*}^\dagger}e^{\hat{\eta}}\ =\ 
e^{\hat{\eta}+\widehat{\theta^*}^\dag}e^{-\theta^*\eta/2}.
\end{equation}
Thus any operator $Q$ has Weyl and anti-normal co-kernels $M_w^Q$ and  $M_{an}^Q$.  \emph{Weyl} and \emph{anti-normal}  symbols $\sigma_w^Q$  and $\sigma_{an}^Q$  are the normal symbols of  $\widehat{M}_w^Q$ and  
$\widehat{M}_{an}^Q$.

By (\ref{eq:BCH}) the  symbols of the same operator $Q$ are related via \emph{Weierstrass transform}
(cp. \textsc{agarwal-wolf}\cite[formulas (5.29), (5.30), (5.31), page 2173]{Agarwal} in a finite-dimensional case, \textsc{dynin}\cite{Dynin-98} and \cite{Dynin-02}) in white noise calculus):
\begin{eqnarray}
\label{eq:wn}
\sigma^Q_w(z^*,z) & = &  e^{-(1/2)\partial_{z^*}\cdot\partial_z}\:\sigma^Q_n(z^*,z),\\
\label{eq:ann}
\sigma^Q_{an}(z^*,z) & = & e^{-N}\:\sigma^Q_n(z^*,z), \\
\label{eq:wa}
\sigma^Q_w(z^*,z) & = &  e^{(1/2)\partial_{z^*}\cdot\partial_z}\:\sigma^Q_{an}(z^*,z),
\end{eqnarray}
where the operator  $e^{\pm(1/2)\partial_{z^*}\cdot\partial_z}$  is the sesqui-linear Fourier transform   of the  multiplication by $e^{\pm(1/2)\zeta^*\zeta}$, i. e, 
\begin{equation}
\label{ }
\partial_{z^*}\cdot\partial_z\ \equiv\ \mbox{Trace}(\partial_{w^*}\partial_z).
\end{equation}

(Note that $e^{\partial_w^*\partial_z}$ is the Fourier transform equivalent   of multiplication by  $e^{\theta^*\eta}$ is a continuous operator on $\mbox{Ent}(\mathcal{C}^{-\infty^*}\times
\mathcal{C}^{-\infty}$, so that  the restrictions to the real diagonal of  $e^{\partial_z^*\partial_z}\sigma(z^*,z)$ are well defined.)
\begin{proposition}
\label{pr:projection}
The anti-normal kernel of an operator $Q=\widehat{M}_{an}$  
\begin{equation}
\langle\ e^{z^*}\ |\ Q\ e^{w}\ \rangle\ =\  e^{z^*w} \sigma_{an}^Q(w^*,w),
\end{equation}
i.e., $Q$ acts on $\Psi(w^*)$ as the  the multiplication  $\sigma_{an}^Q(w^*,w)
\Psi(w^*)\in\mbox{Ent}\big(\mathcal{C}^{\infty})^*\times \mathcal{C}^{\infty}\big)$ followed by the orthogonal projection with the kernel $(e^{z^*w})^*=e^{w^*z}$ onto 
$\mbox{Ent}\big(\mathcal{C}^{\infty})$.
\end{proposition}
\proof
Since
\begin{eqnarray}
& & 
\langle\ e^{z^*}\ | e^{\widehat{\theta^*}^\dagger}e^{\hat{\eta}}\ | e^{w}\ \rangle\  =\
\langle\  e^{\widehat{\theta^*}}e^{z^*}\ |\ e^{\hat{\eta}} e^{w}\ \rangle \\
 & &
=\ \langle\  e^{\theta^*z}e^{z^*}\ |\  e^{z^*\eta}e^{w}\ \rangle\ =\ e^{z^*w}e^{\theta^*w+w^*\eta},
\end{eqnarray}
the anti-normal kernel (\ref{eq:an})
\begin{equation}
\langle\ M_{an}(\theta^*,\eta)\ |\ 
 e^{z^*w}e^{\theta^*w+w^*\eta}\ \rangle\ =\ e^{z^*w}\tilde{M}{an}(w,w^*).
\end{equation}
\qed
\subsection{Number operator}
An operator $Q$ is  a \emph{polynomial operator}  if any its l symbol (and then the other symbols) is a continuous polynomial  on $\mathcal{C}^*\times \mathcal{C}$  It belongs to  belongs to $\mbox{Exp}(\mathcal{C}^{-\infty})$,
and, therefore, their quantizations are tame operators.

The \emph{number operator}  $N\equiv \partial_{z^*}\partial_{z}$  is a polynomial operator with the  quadratic symbol $\sigma_n^N\ =\ z^*z$. Then, by (\ref{eq:ann}) and  (\ref{eq:wn}) the other symbols
\begin{equation}
\label{eq:number}
\sigma^N_{an}(z^*,z)=z^*z+ 1, \quad  \sigma^N_w(z^*,z)= z^*z+1/2.
\end{equation}
The eigenspaces $\mathcal{N}_n,\   n=0,1,2,...,$ of $N$  with the corresponding eigenvalue $n$ is the  space of continuous  homogeneous polynomials of degree ($n$-\emph{bosonic states}). 
 In particular,  the constant  \emph{vacuum state}  $\Psi_0\equiv 1$ corresponds to the eigenvalue $n=0$.

\smallskip
 Gelfand-Schwartz  triple (\ref{eq:Gelfand}) is the topological orthogonal sum of $n$-bosons  triples
\begin{equation}
\label{eq:n}
\mathcal{N}_n^\infty\ \subset\ \mathcal{N}_n^0\ \subset 
\mathcal{N}_n^{-\infty},\quad n=0,1,2,....
\end{equation}

\section{Yang-Mills energy-mass operator} 
\subsection{Energy-mass functional}
In the first order formalism  and the temporal gauge the energy-mass functional (\ref{eq:energy}) is (see, e.g \textsc{faddeev-slavnov} \cite[Section III.2]{Faddeev})
\begin{equation}
\label{eq:H}
H(a,e)\ =\ (1/2)\int_{\mathbb{R}^3}\!d^3x\: \Big((da - [a,a])\cdot (da - [a,a])\ +\
e\cdot e\Big).
\end{equation} 
The functional is invariant under the residual gauge transformations subject to   the temporal restriction $A_0(t,x^k)=0$.
 \begin{proposition}
 \label{pr:non-negative}
Yang-Mills energy-mass functional  of the constrained Cauchy  data is gauge  equivalent to the non-negative functional
\begin{equation}
\label{}
H(a,e)\ =\  (1/2)\int_{\mathbb{R}^3}\!d^3x\: \Big(da\cdot da +[a,a]\cdot  [a,a]\ +\
e\cdot e\Big)
\end{equation} 
\end{proposition}
\proof
 Let $\mathcal{G}^0$ denote the completion of  $\mathcal{G}^\infty$ with respect to the natural $\mathcal{L}^2$-metric on the transformations of 
$\mathcal{S}^0$. Then, by \textsc{dell'antonio-zwanziger} \cite[Proposition 4]{Dell'Antonio},
  \begin{enumerate}
  \item The gauge action of $\mathcal{G}^\infty$  on $\mathcal{A}^\infty\times \mathcal{S}^\infty$  has a unique extension to the continuous action of $\mathcal{G}^0$ on
  \begin{equation}
\label{ }
\mathcal{C}^0\ \equiv\  \mathcal{A}^0\times \mathcal{S}^0.
\end{equation}
  
  \item  The gauge orbits of this action are closures of  $\mathcal{G}^\infty$-orbits.
  
    \item On the euclidean  orbit of every $a$ the Hilbert $\mathcal{L}^2$-norm $\|a^g\|$ attains the absolute minimum  at some gauge  equivalent connection 
    $\breve{a}\in\mathcal{A}^0$. 
   \item Minimizing connections $\breve{a}$ are weakly divergence free:
$\partial^k\breve{a}_k\ =\ 0$. 
\end{enumerate}
Thus the cubic term  in $(da - [a,a])\cdot (da - [a,a])$ vanishes on the minimizing connections. \qed

\subsection{Yang-Mills energy-mass operator and its symbols}
Let the  \emph{quantum Yang-Mills energy-mass operator} $H:\ (\mathcal{C})^\infty
\rightarrow \mbox{Exp}(\mathcal{C}^{-\infty})$  be the anti-normal quantization  of the   energy-mass functional  $H$:
\begin{equation}
\label{}
\sigma^H_{an}(a,e)\ \equiv\ H(a,e)\ =\ H(z^*,z  ),\quad  \zeta=a+ie,\ \zeta^*=a^T-ie^T,
\end{equation}
i.e. $H$ is  the \emph{anti-normal symbol}  of $H$.

\smallskip
The \emph{expectation functional} on non-zero $(\mathcal{C})^\infty\in(\mathcal{C})^\infty$ of a polynomial   operator $Q$ is 
\begin{equation}
\label{ }
\langle Q \rangle (\Psi) \equiv\ \Psi^*Q\Psi/\Psi^*\Psi.
\end{equation}
  
\begin{proposition} \label{pr:geq}
There exists a constant scalar  field $C$ on $\mathbb{R}^3$ such that 
the expectation functional     
\begin{equation}
\label{eq:short}
\langle H\rangle \ \geq \ \langle N\rangle  \  + \ \langle C\rangle,
\end{equation}
where $N$ is the number operator (\ref{eq:number}). 
\end{proposition}

\proof 

(A)\  Let $M$ be the  operator with the non-negative anti-normal symbol 
\begin{equation}
\label{eq:M}
\Omega^M_{an}(z^*,z)\ \equiv\ \int_{\mathbb{R}^3}\!d^3x\:([a,a]\cdot [a,a]+ e\cdot e).
\end{equation}
Then, by Propositions \ref{pr:non-negative} and \ref{pr:projection},
\begin{equation}
\label{eq:A}
\langle H\rangle \ \geq \ \langle M\rangle.
\end{equation}

(B)\ Let  $b_i$ be a basis for $\mbox{Ad}(\mathbb{G})$ with $b_i\cdot b_j=\delta_{ij}$.
Then the structure constants $c^k_{ij}\ =\ [b_i,b_j]\cdot b_k$
 are anti-symmetric under interchanges of  $i,j,k$.  
Thus if   $a=a^ib_i\in\mbox{Ad}(\mathbb{G})$ then 
\begin{equation}
\label{eq:Killing}
a\cdot a\ =\  \mbox{Trace}(a^ta)\ =\ -a^ic^k_{ij}a^lc^j_{kj}\ =\  a^ic^k_{ij}a^lc^k_{lj},
\end{equation}
so that
\begin{eqnarray}
& & 
[a,a]\cdot[a,a]\ =\  a^ia^ja^la^m\, [b_i,b_j]\cdot  [b_l,b_m]\\
& & 
\label{}
=\  a^ia^ja^la^m c^k_{ij}c^k_{lj}\ =\  \sum_k(a^ia^jc^k_{ij})^2.
\end{eqnarray}
 Since $z=(1/\sqrt{2})(a+ie^o)$, the  operator $\partial_z^*\partial_z =  \partial_a^2\ +\ \partial_{e^o}^2$, so that
\begin{equation}
\label{eq:Laplacian}
e^{\partial_z^*\partial_z/2}([a,a]\cdot[a,a])\ \stackrel{(\ref{eq:Killing})}{=}\ [a,a]\cdot[a,a]/2\ +\ a\cdot a.
\end{equation}
Then there is a constant scalar field $C$ such that the Weyl symbol of the operator $M$
\begin{equation}
\label{ }
\sigma^M_w(a,e)\ \stackrel{(\ref{eq:wa}),(\ref{eq:Laplacian})}{=}\ 
\int_{\mathbb{R}^3}\!d^3x\:\big([a,a]\cdot [a,a]/2 \ +\ a\cdot a \ +\ e\cdot e\big)\ +\ 1/2\ +\ C.
\end{equation}
(C)\ The Weyl quantization of  $[a,a]\cdot [a,a]$  is the operator of multiplication with $[a,a]\cdot [a,a]\geq 0$ in the "$(a,e)$-representation" of the canonical commutation relations (cp. \textsc{agarwal-wolf} \cite[Section VII, page 2177]{Agarwal}). In paricular,  its expectation functional is non-negative.

\noindent (D)\ By (\ref{eq:number}),  $\int_{\mathbb{R}^3}\!d^3x\:(a\cdot a \ +\ e\cdot e)\ +\ 1/2$   is the anti-normal symbol of the number operator $N$.

Thus 
\begin{equation}
\label{eq:B}
\langle M \rangle\  \geq \langle N \rangle\ +\ \langle C \rangle.
\end{equation} 
The propostion  follows from  the inequalities (\ref{eq:A}) and (\ref{eq:B}). \qed

\subsection{Bosonic spectrum of Yang-Mills energy-mass operator}
Operfator   $H$ is a non-negative polynomial symmetric operator. It has  a  unique  Friedrichs  extension to an unbounded self-adjoint  operator on the Hilbert space $\mathcal{B}(\mathcal{C}^0)$.  Proposition \ref{pr:geq} implies via the variational mini-max principle (see, e.g.  \textsc{berezin-shubin} \cite[Appendix 2,Proposition 3.2]{Berezin-91}) that its  spectrum is degenerate along with the spectrum of the number operator.

To remove the degeneracy, consider the $n$-particle spaces $\mathcal{N}^\infty_n$ as elementary bosons of spin $n$. Then define the
\emph{bosonic spectrum} of $H$ as  the non-decreasing sequence of its   \emph{spectral values} 
 \begin{equation}
\label{eq:variational}
 \lambda_n(H)\equiv \inf\{\langle H \rangle(\Psi),\ \Psi\in \mathcal{N}^\infty_n\}.
 \end{equation}
 Proposition \ref{pr:geq} implies the main 
\begin{theorem}
\label{th:titular} 
The bosonic spectrum of Yang-Mills energy-mass  operator $H$ is infinite and discrete, i.e. each 
 $\lambda_n(H)$  has a  finite bosonic multiplicity.

The bosonic spectral values  grow at least in  the arithmetical progression:   
\begin{equation}
\label{}
\lambda_n(H)\ \geq\ n\ + \mbox{constant},\quad n=0,\ 1,\ 2, \dots\ .
\end{equation}
\end{theorem}

\end{document}